\documentclass[copyright,creativecommons]{eptcs}
 \usepackage{breakurl}
 \usepackage{epsfig}
\usepackage{amsfonts}
\usepackage{amssymb}
\usepackage{amsmath}
\usepackage{prooftree}

        \providecommand{\event}{Gabriel Ciobanu (Ed.): Membrane Computing and
Biologically Inspired Process Calculi 2009} % adapt
    \providecommand{\volume}{15}   % adapt
    \providecommand{\anno}{2009}   % adapt
    \providecommand{\firstpage}{1} % adapt
    \providecommand{\eid}{1}       % adapt

\newcommand{\qqop}[1]{\mathrel{\makebox[2em]{$#1$}}}

\newcommand{\nat}{\mathbf{N}}
\newcommand{\re}{\mathbf{R}}
\newcommand{\sn}[1]{\langle#1\rangle}
\newcommand{\cn}{^\frown\!\!}
\newcommand{\st}[1]{\widetilde{#1}}
\newcommand{\RR}{\mathcal{R}}
\newcommand{\CC}{\mathcal{C}}
\newcommand{\SC}{\mathcal{SC}}
\newcommand{\ltrans}[1]{\xrightarrow{#1}}
\newcommand{\lmap}[2]{\xrightarrow[#1]{#2}}
\newcommand{\set}[1]{\{#1\}}
\newcommand{\type}[1]{{\it type}(#1)}
\newcommand{\stype}[1]{{\it stype}(#1)}
\newcommand{\seqts}{sequence\ types}
\newcommand{\seqt}{sequence\ type}
\newcommand{\mUnion}{\uplus}
\newcommand{\EE}{\mathcal{E}}
\newcommand{\G}{\Gamma}
\newcommand{\Var}{V}
\newcommand{\genVar}{\chi}

\newcommand{\g}{\phi}

\newcommand{\SALTA}[1]{}

\newcommand{\lis}[2]{\langle#1,\ldots,#2\rangle}

\newcommand{\PP}{\mathcal{P}}
\newcommand{\Seq}{\mathcal{S}}
\newcommand{\XX}{\mathcal{X}}
\newcommand{\TV}{\mathcal{TV}}
\newcommand{\SV}{\mathcal{SV}}
\newcommand{\TT}{\mathcal{T}}

\newcommand{\SSq}{\mathcal{S}}

\newcommand{\VV}{\mathcal{V}}
\newcommand{\xx}{\widetilde{x}}
\newcommand{\yy}{\widetilde{y}}
\newcommand{\zz}{\widetilde{z}}

\newcommand{\Ltrans}[1]{\Longrightarrow}
\newcommand{\Loop}[1]{\left(#1\right)^L}

\newcommand{\into}{\ensuremath{\,\rfloor}\,}
\newcommand{\pipe}{\ensuremath{\,|\,}}
\newcommand{\phole}{\square}

\newcommand{\agr}{\quad\big|\quad}

\newcommand{\mycdot}{\!\cdot\!}

\newtheorem{theorem}{Theorem}[section]
\newtheorem{definition}[theorem]{Definition}

\title{A Type System for a Stochastic CLS\thanks{This work was partly funded by the project BioBIT of the Regione
Piemonte.}}
\author{Mariangiola Dezani-Ciancaglini
\institute{Dipartimento di Informatica, Universit\`a di Torino}
\email{dezani@di.unito.it}
\and
Paola Giannini
\institute{Dipartimento di Informatica, Universit\`a del Piemonte Orientale}
\email{giannini@mfn.unipmn.it}
\and
Angelo Troina
\institute{Dipartimento di Informatica, Universit\`a di Torino}
\email{troina@di.unito.it}
}

\def\titlerunning{A Type System for a Stochastic CLS}
\def\authorrunning{M. Dezani-Ciancaglini, P. Giannini and A. Troina}

\begin{document}
\maketitle

\begin{abstract}
The Stochastic Calculus of Looping Sequences is suitable to describe
the evolution of microbiological systems, taking into account the
speed of the described activities. We propose a type system for this
calculus that models how the presence of positive and negative catalysers
can modify these speeds. We claim that types are the right
abstraction in order to represent the interaction between elements
without specifying exactly the element positions. Our claim is
supported through an example modelling the lactose operon.
\end{abstract}

\section{Introduction}
The Calculus of Looping Sequences (CLS for
short)~\cite{BMMT06,BMMT06s,M07}, is a formalism for describing
biological systems and their evolution. CLS is based on term
rewriting, given a set of predefined rules modelling the activities
one would like to describe. The model has been extended with several
features, such as a commutative parallel composition operator, and
some semantic means, such as bisimulations \cite{BMMT06s,BMMT08},
which are common in process calculi. This permits to combine the
simplicity of notation of rewrite systems with the advantage of a
form of compositionality. A Stochastic version of CLS (SCLS for
short) is proposed in~\cite{BMMTT08}. Rates are associated with
rewrite rules in order to model the speed of the described
activities. Therefore, transitions derived in SCLS are driven by a
rate that models the parameter of an exponential distribution and
characterizes the stochastic behaviour of the transition. The choice
of the next rule to be applied and of the time of its application is
based on the classical Gillespie's algorithm~\cite{G77}.

Defining a stochastic semantics for CLS  requires a correct
enumeration of all the possible and distinct ways to apply each
rewrite rule within a term. A single pattern may have several,
though isomorphic, matches within a CLS term. In this paper, we
simplify the counting mechanism used in~\cite{BMMTT08} by imposing
some restrictions on the patterns modelling the rewrite rules. Each
rewrite rule states explicitly the types of the elements whose
occurrence are able to speed-up or slow-down a reaction. The
occurrences of the elements of these types are then processed by a
rate function (instead of a rate constant) which is used to compute
the actual rate of a transition. We show how we can define patterns
in our typed stochastic framework to model some common biological
activities, and, in particular, we underline the possibility to
combine the modelling of positive and negative catalysers within a single
rule by reproducing a general case of osmosis.

Finally, as a complete modelling application, we show the
expressiveness of our formalism by describing the lactose operon in
{\em Escherichia Coli}.

\paragraph{Summary}
The remainder of this paper is organized as follows. In
Section~\ref{CLS_formalism} we formally recall the Calculus of
Looping Sequence. In Section~\ref{SCLS_types} we introduce our typed
stochastic extension and we give some guidelines for the modelling
of biological systems. In Sections~\ref{secApplication} we use our
framework to model the lactose operon of {\em Escherichia Coli}.
Finally, in Section~\ref{secRelatedWorks} we draw our conclusions
and we present some related work.

\section{The Calculus of Looping Sequences}\label{CLS_formalism}

In this section we recall the Calculus of Looping Sequences (CLS).
CLS is essentially based on term rewriting, hence a CLS model
consists of a term and a set of rewrite rules. The term is intended
to represent the structure of the modelled system, and the rewrite
rules to represent the events that may cause the system to evolve.

We start with defining the syntax of terms. We assume a possibly
infinite alphabet $\EE$ of symbols ranged over by $a,b,c,\ldots$.

\begin{definition}[Terms] \emph{Terms} $T$ and \emph{sequences} $S$ of {\em CLS} are
given by the following grammar:
\[
\begin{array}{lcl}
T\; & \qqop{::=} &S \agr \Loop{S} \into T \agr T \pipe T\\
S\; & \qqop{::=}& \epsilon \agr a  \agr S \cdot S
\end{array}
\]
 where $a$ is a generic element of $\EE$, and $\epsilon$ represents
the empty sequence. We denote with $\TT$ the infinite set of terms,
and with $\Seq$ the infinite set of sequences.
\end{definition}

In CLS we have a sequencing operator $\_\cdot\_$, a looping operator
$\Loop{\_}$, a parallel composition operator $\_\pipe\_$ and a
containment operator $\_\into\_$. Sequencing can be used to
concatenate elements of the alphabet $\EE$. The empty sequence
$\epsilon$ denotes the concatenation of zero symbols. A term can be
either a sequence or a looping sequence (that is the application of
the looping operator to a sequence) containing another term, or the
parallel composition of two terms.
By definition, looping and containment are always applied together,
hence we can consider them as a single binary operator $\Loop{\_}
\into \_$ which applies to one sequence and one term.

The biological interpretation of the operators is the following: the
main entities which occur in cells are DNA and RNA strands,
proteins, membranes, and other macro--molecules. DNA strands (and
similarly RNA strands) are sequences of nucleic acids, but they can
be seen also at a higher level of abstraction as sequences of genes.
Proteins are sequence of amino acids which usually have a very
complex three--dimensional structure. In a protein there are usually
(relatively) few subsequences, called domains, which actually are
able to interact with other entities by means of chemical reactions.
CLS sequences can model DNA/RNA strands and proteins by describing
each gene or each domain with a symbol of the alphabet. Membranes
are closed surfaces, often interspersed with proteins, which may
contain something. A closed surface can be modelled by a looping
sequence. The elements (or the subsequences) of the looping sequence
may represent the proteins on the membrane, and by the containment
operator it is possible to specify the content of the membrane.
Other macro--molecules can be modelled as single alphabet symbols, or
as short sequences. Finally, juxtaposition of entities can be
described by the parallel composition of their representations.

\begin{figure}[t]
\begin{center}
\begin{minipage}{0.98\textwidth}
\begin{center}
\psfig{figure=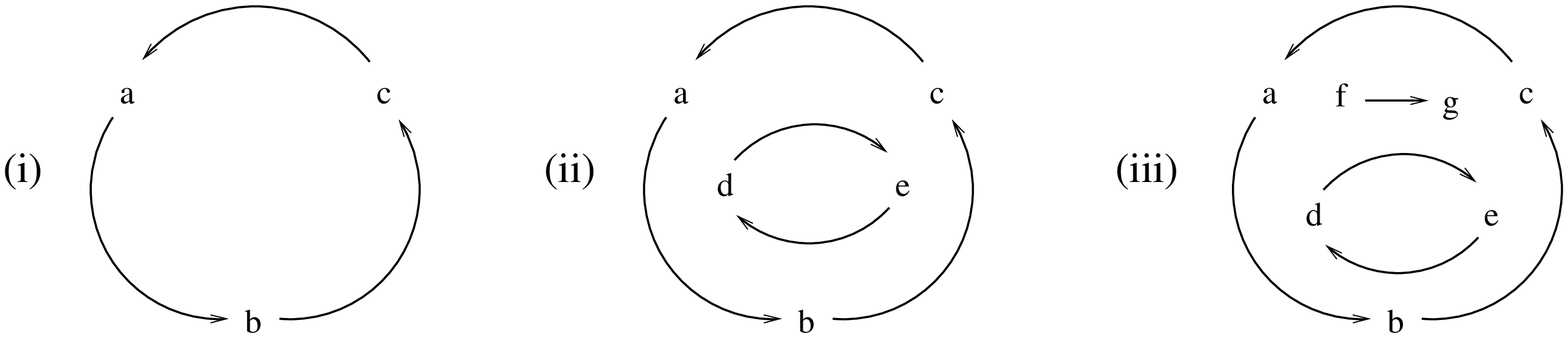, scale=0.43}
\end{center}
\caption{(i) represents $\Loop{a
\cdot b \cdot c}$; (ii) represents $\Loop{a
\cdot b \cdot c} \into \Loop{d \cdot e}$; (iii) represents $\Loop{a \cdot b
\cdot c} \into (\Loop{d \cdot e} \pipe f \cdot g)$.}
\label{fig:loop_seq_fig_CLS}
\end{minipage}
\end{center}
\end{figure}

Brackets can be used to indicate the order of application of the
operators, and we assume $\Loop{\_} \into \_$ to have precedence
over $\_\pipe\_$. In Figure~\ref{fig:loop_seq_fig_CLS} we show some
examples of CLS terms and their visual representation, using $\Loop{S}$ as a short-cut for $\Loop{S} \into \epsilon$.

In CLS we may have syntactically different terms representing the
same structure. We introduce a structural congruence relation to
identify such terms.

\begin{definition}[Structural Congruence] The structural
congruence relations $\equiv_S$ and $\equiv_T$ are the least
congruence relations on sequences and on terms, respectively,
satisfying the following rules:
\[
\begin{array}{c}
S_1 \cdot ( S_2 \cdot S_3 )
    \equiv_S ( S_1 \cdot S_2 ) \cdot S_3 \qquad
S \cdot \epsilon \equiv_S \epsilon \cdot S
    \equiv_S S
\\
S_1 \equiv_S S_2\ \mbox{ implies }\ S_1 \equiv_T S_2\ \mbox{ and }\
    \Loop{S_1} \into T \equiv_T \Loop{S_2} \into T\\
T_1 \pipe T_2 \equiv_T T_2 \pipe T_1 \qquad T_1 \pipe ( T_2 \pipe
T_3 ) \equiv_T (T_1 \pipe T_2) \pipe T_3 \qquad
T \pipe \epsilon \equiv_T T \\
\Loop{\epsilon} \into \epsilon \equiv_T \epsilon \qquad \Loop{S_1
\cdot S_2} \into T \equiv_T \Loop{S_2 \cdot S_1} \into T
\end{array}
\]
\end{definition}

Rules of the structural congruence state the associativity of
$\cdot$ and $\pipe$, the commutativity of the latter and the neutral
role of $\epsilon$. Moreover, axiom $\Loop{S_1 \cdot S_2} \into T
\equiv_T \Loop{S_2 \cdot S_1} \into T$ says that looping sequences
can rotate. In the following, for simplicity, we will use $\equiv$
in place of $\equiv_T$.

Rewrite rules will be defined essentially as pairs of terms, with
the first term describing the portion of the system in which the
event modelled by the rule may occur, and the second term describing
how that portion of the system changes when the event occurs. In the
terms of a rewrite rule we allow the use of variables. As a
consequence, a rule will be applicable to all terms which can be
obtained by properly instantiating its variables. Variables can be
of three kinds: two of these are associated with the two different
syntactic categories of terms and sequences, and one is associated
with single alphabet elements. We assume a set of term variables
$\TV$ ranged over by $X,Y,Z,\ldots$, a set of sequence variables $\SV$
ranged over by $\xx,\yy,\zz,\ldots$, and a set of element variables
$\XX$ ranged over by $x,y,z,\ldots$. All these sets are possibly
infinite and pairwise disjoint. We denote by $\VV$ the set of all
variables, $\VV = \TV \cup \SV \cup \XX$, and with $\genVar$ a generic
variable of $\VV$. Hence, a pattern is a term that may include
variables.

\begin{definition}[Patterns] \emph{Patterns} $P$ and \emph{sequence patterns}
$SP$ of {\em CLS} are given by the following grammar:
\[
\begin{array}{lcl}
P\; & \qqop{::=} SP \agr \Loop{SP} \into P \agr P \pipe P \agr X\\
SP\; & \qqop{::=} \epsilon \agr a  \agr SP \cdot SP \agr \xx \agr x
\end{array}
\]
where $a$ is a generic element of $\EE$, and $X,\xx$ and $x$ are
generic elements of $\TV,\SV$ and $\XX$, respectively. We denote with
$\PP$ the infinite set of patterns.
\end{definition}

We assume the structural congruence relation to be trivially
extended to patterns. An \emph{instantiation} is a partial function
$\sigma : \VV \rightarrow \TT$. An instantiation must preserve the
type of variables, thus for $X \in \TV, \xx \in \SV$ and $x \in \XX$
we have $\sigma(X) \in \TT, \sigma(\xx) \in \Seq$ and $\sigma(x) \in
\EE$, respectively. Given $P \in \PP$, with $P \sigma$ we denote the
term obtained by replacing each occurrence of each variable $\genVar
\in \VV$ appearing in $P$ with the corresponding term
$\sigma(\genVar)$. With $\Sigma$ we denote the set of all the possible
instantiations and, given $P \in \PP$, with $Var(P)$ we denote the
set of variables appearing in $P$. Now we define rewrite rules.

\begin{definition}[Rewrite Rules] A rewrite rule is a pair of patterns
$(P_1,P_2)$, denoted with $P_1 \! \mapsto \! P_2$, where $P_1,P_2
\in \PP$, $P_1 \not\equiv \epsilon$ and such that $Var(P_2)
\subseteq Var(P_1)$.
\end{definition}

A rewrite rule $P_1 \! \mapsto \! P_2$ states that a term $P_1
\sigma$, obtained by instantiating variables in $P_1$ by some
instantiation function $\sigma$, can be transformed into the term
$P_2\sigma$. We define the semantics of CLS as a transition system,
in which states correspond to terms, and transitions correspond to
rule applications.

We define the semantics of CLS by resorting to the notion of
contexts.

\begin{definition}[Contexts] \label{context} \emph{Contexts} $C$ are defined as:
\[
C ::= \phole \agr C \pipe T \agr T \pipe C \agr \Loop{S} \into C
\]
where $T \in \TT$ and $S \in \SSq$. The context $\phole$ is called
the \emph{empty context}. We denote with $\CC$ the infinite set of
contexts.
\end{definition}

By definition, every context contains a single hole $\phole$. Let us
assume $C\in \CC$, with $C[T]$ we denote the term obtained by
replacing $\phole$ with $T$ in $C$. The structural equivalence is extended to
contexts in the natural way (i.e. by considering $\phole$ as a new
and unique symbol of the alphabet $\EE$).

Rewrite rules can be applied to terms only if they occur in a legal
context. Note that the general form of rewrite rules does not permit
to have sequences as contexts. A rewrite rule introducing a parallel
composition on the right hand side (as $a \mapsto b \pipe c$)
applied to an element of a sequence (e.g., $m \mycdot a \mycdot m$)
would result into a syntactically incorrect term (in this case $m
\cdot (b\pipe c) \cdot m$). To modify a sequence, a pattern
representing the whole sequence must appear in the rule. For
example, rule $a \mycdot \xx \mapsto a \pipe \xx$ can be applied to
any sequence starting with element $a$, and, hence, the term
$a\mycdot b$ can be rewritten as $a \pipe b$, and the term $a
\mycdot b \mycdot c$ can be rewritten as $a \pipe b \mycdot c$.

The semantics of CLS is defined as follows.

\begin{definition}[Semantics]
Given a finite set of rewrite rules $\RR$, the \emph{semantics} of
{\em CLS} is the least relation closed with respect to $\equiv$ and
satisfying the following rule:
\[
\prooftree
\begin{array}{c} P_1 \mapsto P_2 \in \RR \quad \sigma \in \Sigma \quad P_1\sigma \not\equiv
\epsilon  \quad C\in \CC \end{array}
\justifies C[P_1\sigma] \ltrans{} C[P_2\sigma]
\endprooftree
\]
\end{definition}

As usual we denote with $\ltrans{}^*$ the reflexive and transitive
closure of $\ltrans{}$.

Given a set of rewrite rules $\RR$, the behaviour of a term $T$ is
the tree of terms to which  $T$ may reduce. Thus, a \emph{model} in
CLS is given by a term describing the initial state of the system
and by a set of rewrite rules describing all the events that may
occur.

\section{Typed Stochastic CLS}\label{SCLS_types}

In this section we show how types are used to enhance the
expressivity of CLS. In particular, we use types to focus on
quantitative aspects of CLS, by showing how to model the speeds of
the biological activities.

We classify elements in $\EE$ with types. Intuitively, given a
molecule represented by an element $a$ in $\EE$, we associate a type to it
which specifies the kind of the molecule. For an element $a$, we distinguish
between occurrences of a single $a$ in parallel with other terms, for which we
use a {\em basic type $t$}, and occurrences of $a$ within a sequence, for
which we use a {\em \seqt\ $\st{t}$}. So, types specify the kind of
elements and their positioning. In the following, with type, we
mean either a basic type or a sequence type and we use $t$ to range over both basic and sequence types. The metavariable $\tau$
ranges over multi-sets of types. By $t\in_n\tau$ we denote that type
$t$ occurs $n$ times in $\tau$, and $\mUnion$ is the union on
multisets.

Let $\G$ be a type assignment such that for $a\in\EE$, if $\G(a)=t$,
then $t$ and $\st{t}$ are the types for $a$. The type of a term (or
sequence) is the multiset of types (or \seqts) of its outermost
component. This is formalised in the following definition of {\it type} of a term and {\it stype} of a sequence.
\begin{definition}[Mappings {\it type} and {\it stype}]
The mappings {\it type} and {\it stype} are defined by induction on terms and sequences as follows:
\begin{itemize}
 \item
 \begin{itemize}
   \item $\type{ \Loop{S} \into T}=\stype{ S}$
  \item $\type{T_1 \pipe T_2}=\type{T_1}\mUnion\type{T_2}$
   \item $\type{S_1 \cdot S_2}=\stype{S_1 \cdot S_2}$
    \item $\type{a}=\set{\G(a)}$
 \end{itemize}
 \item
 \begin{itemize}
   \item $\stype{S_1 \cdot S_2}=\stype{S_1}\mUnion\stype{S_2}$
  \item $\stype{a}=\set{\st{\G(a)}}$
\end{itemize}
 \end{itemize}
\end{definition}
For example if $\G(a)=t_a$, $\G(b)=t_b$ and $\G(c)=t_c$ we have $\type{a\pipe a\pipe c}=\set{t_a,t_a,t_c}$, $\type{b\cdot c\cdot c}=\set{\st{t_b},\st{t_c},\st{t_c}}$, $\type{a\pipe a\pipe c\pipe \Loop{b\cdot c\cdot c}\into a}=\set{t_a,t_a,t_c,\st{t_b},\st{t_c},\st{t_c}}$ and $\type{\Loop{b\cdot c\cdot c}\into {(a\pipe a\pipe a\pipe c)}}=\set{\st{t_b},\st{t_c},\st{t_c}}$.
\bigskip

Term transitions are labelled with a {\it rate} $r$, a real number,
$T \ltrans{r} T'$, modelling the speed of the transition. The number
$r$ depends on the types and multiplicity of the elements
interacting.

To compute the rate of transitions we associate to each rule, $P \!
\mapsto \! P'$ the information which is relevant to the application
of the rule. This is expressed by giving:
\begin{itemize}
  \item for each variable $\genVar$ in the pattern $P$, the types of the
  elements that influence the speed of the application of the rule,
   \item a weighting function that combines the multiplicity of
  types
 on single variables, producing the final rate.
\end{itemize}
We provide this information as follows. Given a pattern $P$, let
$\Var(P)=\lis{\genVar_1}{\genVar_m}$ be the list of (sequence, term,
and element) variables of $P$ in left-to-right order of occurrence.
\begin{itemize}
  \item To each $\genVar_i$ we associate a list $\Pi_{i}=\lis{t^{(i)}_1} {t^{(i)}_{p_i}}$ of types,
  \item Moreover, let  $\g:\nat^q\to\re$ be a function from a list of $q=\sum_{1\leq i\leq m}p_i$ integers to a real.
\end{itemize}
The rewrite rules of our {\em Typed Stochastic} CLS  (TSCLS for short) are of the  shape
 \[
P \lmap{\g}{\overline{\Pi}}{} P'
 \]
 where $\overline{\Pi}=\lis{\Pi_1}{\Pi_m}$.

 For example as discussed in the following subsection the transformation of the element $a$ into the element $b$ inhibited by the presence of the element $c$ can be described by the rule
\begin{equation}\label{eq:rule}
a\pipe X \lmap{\g}{\sn{\sn{t_a,t_c}}}{} b\pipe X
\end{equation}
 where $\g=\lambda n_1n_2.\frac{(n_1+1)\times k}{\text{if } n_2=0 \text{ then } 1 \text{ else } n_2\times k'}$, and $k,k'$ are the kinetic constant of the state change of $a$ into $b$ and the deceleration due to the presence of one inhibitor $c$, respectively.

We consider local interactions, that is interactions between
elements in the same compartment. When applying a rule, to take into
account a whole compartment, we redefine the notion of context
by enforcing the property that the hole of a context embraces a whole
compartment as follows:
\begin{definition}[Stochastic Contexts] \emph{Stochastic Contexts} $C$ are defined as:
\[
C ::= \phole  \agr T \pipe \Loop{S} \into C
\]
where $T \in \TT$ and $S \in \SSq$. We denote with $\SC$ the infinite set of
stochastic contexts.
\end{definition}
We can now define the typed semantics.

\begin{definition}[Typed Stochastic Semantics]
Given a finite set of rewrite rules $\RR$, the \emph{semantics} of
{\em TSCLS} is the least relation closed with respect to $\equiv$ and
satisfying the following rule:
\[
\prooftree
\begin{array}{c} P_1
\lmap{\g}{\overline{\Pi}}{} P_2\in \RR \quad \sigma \in \Sigma \quad P_1\sigma \not\equiv
\epsilon  \quad C\in \SC\\
\type{\sigma(\genVar_i)}=\tau_i\quad\quad
t^{(i)}_{j}\in_{n^{(i)}_j}\tau_i\quad(1\leq j\leq p_i)\quad(1\leq i\leq m)\\
r=\g(n^{(1)}_1,\ldots
,n^{(1)}_{p_1},\cdots,n^{(m)}_1,\ldots , n^{(m)}_{p_m})\\$\;$
\end{array} \justifies C[P_1\sigma] \ltrans{r} C[P_2\sigma]
\endprooftree
\]
\end{definition}
For example, applying rule (\ref{eq:rule}) with the empty context to
the term $a\pipe a\pipe c$ we have:
\[a\pipe a\pipe c\ltrans{\frac{2\times k}{1\times k'}}a\pipe b\pipe c\ltrans{\frac{1\times k}{1\times k'}}b\pipe b\pipe c\]
and to the term $a\pipe a\pipe c\pipe \Loop{b\cdot c\cdot c}\into a$
we have:
\[a\pipe a\pipe c\pipe \Loop{b\cdot c\cdot c}\into a\ltrans{\frac{2\times k}{1\times k'}}a\pipe b\pipe c\pipe \Loop{b\cdot c\cdot c}\into a\ltrans{\frac{1\times k}{1\times k'}}b\pipe b\pipe c\pipe \Loop{b\cdot c\cdot c}\into a\]
Applying (\ref{eq:rule}) with the context $\epsilon\pipe\Loop{b\cdot
c\cdot c}\into \phole$ to the term $\Loop{b\cdot c\cdot c}\into
{a\pipe a\pipe a\pipe c}$ we get:
\[\Loop{b\cdot c\cdot c}\into {a\pipe a\pipe a\pipe c}\ltrans{\frac{3\times k}{1\times k'}}\Loop{b\cdot c\cdot c}\into {a\pipe a\pipe b\pipe c}\ltrans{\frac{2\times k}{1\times k'}}\Loop{b\cdot c\cdot c}\into {a\pipe b\pipe b\pipe c}\ltrans{\frac{1\times k}{1\times k'}}\Loop{b\cdot c\cdot c}\into {b\pipe b\pipe b\pipe
c}\]
Note that we cannot use Definition \ref{context} for contexts,
since we would not count correctly the numbers of elements which
influence the speed of transformations. For example, again rule
(\ref{eq:rule}) applied to the term $a\pipe a\pipe c$ with the
context $\phole\pipe a\pipe c$ would produce the wrong transition:
\[a\pipe a\pipe c\ltrans{k}a\pipe b\pipe c.\]

Given the Continuous Time Markov Chain (CTMC) obtained from the
transition system resulting from our typed stochastic semantics, we
can follow a standard simulation procedure. Roughly speaking, the
algorithm starts from the initial term (representing a state of the
CTMC) and performs a sequence of steps by moving from state to
state. At each step a global clock variable (initially set to zero)
is incremented by a random quantity which is exponentially
distributed with the exit rate of the current state as parameter,
and the next state is randomly chosen with a probability
proportional to the rates of the exit transitions.

The \emph{race condition} described above implements the fact that, on the lines of
Gillespie's algorithm~\cite{G77}, when different reactions are competing with different
rates, the ones which are not chosen should restart the competition at the following
step.

\subsection{Modelling Guidelines}
In the remain of this section we will put at work the TSCLS calculus
in order to model biomolecular events of interest.

\begin{itemize}
\item
The application rate in the case of the {\em change of state of an
elementary object} is proportional to the  number of objects which
are present. For this reason if $t_a$ is the type of the object $a$
and $k$ is the kinetic constant of the state change of $a$ into $b$
we can describe this chemical reaction by the following rewrite
rule:
\[
a\pipe X
\lmap{\g}{\sn{\sn {t_a}}}{} b\pipe X\]
where $\g=\lambda n. (n+1)\times k$. Using this rule we get for example:
\[\Loop{m} \into (a\pipe a\pipe a)\ltrans{3k}\Loop{m} \into (b\pipe a\pipe a)\]
\[\Loop{m} \into (a\pipe a\cdot a)\ltrans{k}\Loop{m} \into (b\pipe a\cdot a)\]
where $m$ is any membrane.

\item In the process of {\em complexation}, two elementary objects in the same
compartment are combined to produce a new object. The application
rate is then proportional to the product of the numbers of
occurrences of the two objects. Assuming that $t_a$ and $t_b$ are
the types of $a$ and $b$ we get:
\[
a\pipe b\pipe X \lmap{\g}{\sn{\sn {t_a,t_b}}}{} c\pipe X\] where
$\g=\lambda n_1n_2.(n_1+1)\times(n_2+1)\times k$ and $k$ is the
kinetic constant of the modelled chemical reaction.

Using the same conventions a similar and simpler rule describes {\em decomplexation}:
\[
c\pipe X
\lmap{\g}{\sn{\sn {t_c}}}{} a\pipe b\pipe X\]
where $\g=\lambda n. (n+1)\times k$.

\item Another phenomenon which can be easily rendered in our
formalism is the {\em osmosis} regulating the quantity of water
inside and outside a cell for a dilute solution of non-dissociating
substances. In fact in this case according to \cite{TZ06} the total
flow is $L_p\frac{S}{V}\Delta\psi_w$, where $L_p$ is the hydraulic
conductivity constant, which depends on the semi-permeability
properties of the membrane, $S$ is the surface of the cell, $V$ is
the volume of the cell, $\Delta\psi_w=\psi_{w(ext)}-\psi_{w(int)}$
is the difference between the water potentials outside and inside
the cell. The water potential for non-dissociating substances is the
sum of the solute potential $\psi_s=-{\sf RT}c_s$ (where $\sf R$ is
the gas constant, $\sf T$ is the absolute temperature and $c_s$ is
the solute concentration) and the pressure potential $\psi_p$ (which
depends on the elastic properties of the membrane  and on the cell
wall). We can therefore consider the rate of flow of water
proportional (via a constant $k$) to
$\frac{S}{V}(c_{s(ext)}-c_{s(int)}$), where the sign of this real
gives the direction of the flow. The membrane crossing of the
element $a$ according to the concentration of the elements $b$
inside and outside the cell is given by the pairs of rules:
\[\begin{array}{c}
\Loop{\xx} \into
(X\pipe a)\pipe Y\lmap{\g}{\sn{\sn{},\sn{t_a,t_b},\sn{t_a,t_b}}}\Loop{\xx}
\into X\pipe a\pipe Y\\
\Loop{\xx} \into X\pipe
a\pipe Y\lmap{\g'}{\sn{\sn{},\sn{t_a,t_b},\sn{t_a,t_b}}}\Loop{\xx} \into
(X\pipe a)\pipe Y
\end{array}\]
where $\begin{array}{l}\g=\lambda n_1n_2n_3n_4.\frac{S}{V}\times(\frac{n_2}{(n_1+1)V_a+n_2V_b}-\frac{n_4}{(n_3+1)V_a+n_4V_b})\times k\\ \g'=\lambda n_1n_2n_3n_4.\frac{S}{V}\times(\frac{n_4}{(n_3+1)V_a+n_4V_b}-\frac{n_2}{(n_1+1)V_a+n_2V_b})\times k\end{array}$ and $V_a,V_b$ are the volumes of the elements $a$ and $b$, respectively.

The {\em positive catalysis} of osmosis by the presence of elements $c$ on the membrane is rendered by:
\[\begin{array}{c}
\Loop{\xx} \into
(X\pipe a)\pipe Y\lmap{\g}{\sn{\sn{\st{t_c}},\sn{t_a,t_b},\sn{t_a,t_b}}}\Loop{\xx}
\into X\pipe a\pipe Y\\
\Loop{\xx} \into X\pipe
a\pipe Y\lmap{\g'}{\sn{\sn{\st{t_c}},\sn{t_a,t_b},\sn{t_a,t_b}}}\Loop{\xx} \into
(X\pipe a)\pipe Y
\end{array}\]
where $\begin{array}{l}\g=\lambda n_1n_2n_3n_4n_5.(n_1\times k_c+1)\times\frac{S}{V}\times(\frac{n_3}{(n_2+1)V_a+n_3V_b}-\frac{n_5}{(n_4+1)V_a+n_5V_b})\times k\\ \g'=\lambda n_1n_2n_3n_4n_5.(n_1\times k_c+1)\times\frac{S}{V}\times(\frac{n_5}{(n_4+1)V_a+n_5V_b}-\frac{n_3}{(n_2+1)V_a+n_3V_b})\times k\end{array}$ and $k_c$ is the acceleration due to the presence of one element $c$.

Similarly the {\em inhibition} of osmosis by the presence of elements $c$ on the membrane is rendered by:
\[\begin{array}{c}
\Loop{\xx} \into
(X\pipe a)\pipe Y\lmap{\g}{\sn{\sn{\st{t_c}},\sn{t_a,t_b},\sn{t_a,t_b}}}\Loop{\xx}
\into X\pipe a\pipe Y\\
\Loop{\xx} \into X\pipe
a\pipe Y\lmap{\g'}{\sn{\sn{\st{t_c}},\sn{t_a,t_b},\sn{t_a,t_b}}}\Loop{\xx} \into
(X\pipe a)\pipe Y
\end{array}\]
where $\begin{array}{l}\g=\lambda n_1n_2n_3n_4n_5.\frac{1}{\text{if } n_1=0 \text{ then } 1 \text{ else } n_1\times k_c}\times\frac{S}{V}\times(\frac{n_3}{(n_2+1)V_a+n_3V_b}-\frac{n_5}{(n_4+1)V_a+n_5V_b})\times k\\ \g'=\lambda n_1n_2n_3n_4n_5.\frac{1}{\text{if } n_1=0 \text{ then } 1 \text{ else } n_1\times k_c}\times\frac{S}{V}\times(\frac{n_5}{(n_4+1)V_a+n_5V_b}-\frac{n_3}{(n_2+1)V_a+n_3V_b})\times k\end{array}$ and $k_c$ is the deceleration due to the presence of one element $c$.

\item
If the rule
\[P_1
\lmap{\g}{\overline{\Pi}}{} P_2\] describes an event, in order to
express that this event is {\em positively catalysed} by an element $c$ we can
modify the rewrite rule as follows.

If $P_1\equiv P_1'\pipe X$, the type list of $X$ is $\Pi_X$ and the
weighting function $\g$ is $\lambda\overline n\overline{n_X}.e$,
where $\overline n$ takes into account the elements occurring in
$P'_1$ and $\overline{n_X}$ takes into account the elements
occurring in $X$, we define:
\begin{itemize}
\item $\Pi_X'$ as the list whose head is $t_c$ and whose tail is $\Pi_X$,
\item $\g'=\lambda\overline n n_c\overline{n_X}.e\times (n_c\times k+1)$,
\end{itemize}
where $k$ is the acceleration due to the presence of one positive catalyser
$c$. The new rule is obtained from the old one by replacing $\Pi_X'$
and $\g'$ to $\Pi_X$ and $\g$, respectively.

Otherwise if $P_1\not\equiv P_1'\pipe X$, the new rule is:
\[P_1\pipe X
\lmap{\g'}{\overline{\Pi}\cn\sn{\sn{t_c}}}{} P_2\pipe X\] where
$\cn\;$ represents list concatenation and if $\g=\lambda\overline
n.e$, then $\g'=\lambda\overline n n_c.e\times (n_c\times k+1)$.

Similarly we can represent the effect of an {\em inhibitor} just
replacing the inserted multiplications by divisions. We can also
represent in one rule both positive and negative catalysers. For example to
add the effect of a positive catalyser $c$ and an inhibitor $d$ to the rule
$P_1\lmap{\g}{\overline{\Pi}}{} P_2$ if $P_1\equiv P_1'\pipe X$ and
$\Pi_X$, $\g$ are as above we define:
\begin{itemize}
\item $\Pi_X'=\sn{t_c,t_d}\cn\,\Pi_X$,
\item $\g'=\lambda\overline n n_cn_d\overline{n_X}.e\times \frac{n_c\times k+1}{\text{if } n_d=0 \text{ then } 1 \text{ else } n_d\times k'}$,
\end{itemize}
where $k$ is the acceleration due to the presence of one positive catalyser
$c$ and $k'$ is the deceleration due to the presence of one
inhibitor $d$.

Otherwise if $P_1\not\equiv P_1'\pipe X$, the new rule is:
\[P_1\pipe X
\lmap{\g'}{\overline{\Pi}\cn\sn{\sn{t_c,t_d}}}{} P_2\pipe X\]
where if $\g=\lambda\overline n.e$, then $\g'=\lambda\overline n n_cn_d.e\frac{n_c\times k+1}{\text{if } n_d=0 \text{ then } 1 \text{ else } n_d\times k'}$.

\end{itemize}
Looking at the previous examples, we claim that our formalism
enlightens better than other formalisms the duality between the
roles of positive and negative catalysers.

\section{An Application: The Lactose Operon}\label{secApplication}
To show that our framework can be easily used to model and simulate cellular
pathways, we give a model of the well-known regulation process
of the lactose operon in {\em Escherichia coli}.

E. coli is a bacterium often present in the intestine of many
animals. It is one of the most completely studied of all living
things and it is a favourite organism for genetic engineering.
Cultures of E. coli can be made to produce unlimited quantities of
the product of an introduced gene. As most bacteria, E.coli is often
exposed to a constantly changing physical and chemical environment,
and reacts to changes in its environment through changes in the
kinds of enzymes it produces. In order to save energy, bacteria do
not synthesize degradative enzymes unless the substrates for these
enzymes are present in the environment. For example, E. coli does
not synthesize the enzymes that degrade lactose unless lactose is in
the environment. This result is obtained by controlling the
transcription of some genes into the corresponding enzymes.

Two enzymes are involved in lactose degradation: the \emph{lactose
permease}, which is incorporated in the membrane of the bacterium
and actively transports the sugar into the cell, and the \emph{beta
galactosidase}, which splits lactose into glucose and galactose. The
bacterium produces also the \emph{transacetylase} enzyme, whose role
in the lactose degradation is marginal.

The sequence of genes in the DNA of E. coli which produces the
described enzymes, is known as the \emph{lactose operon}.

\begin{figure}[t]
\begin{center}
\psfig{figure=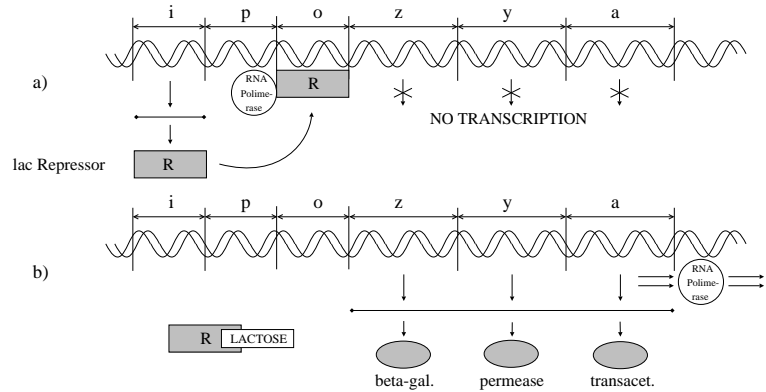,scale=0.12}
\end{center}
\caption{The regulation process in the Lac Operon.}
\label{fig:lac-operon-combo}
\end{figure}

The first three genes of the operon (i, p and o) regulate the
production of the enzymes, and the last three (z, y and a), called
\emph{structural genes}, are transcribed (when allowed) into the
mRNA for beta galactosidase, lactose permease and transacetylase,
respectively.

The regulation process is as follows (see
Figure~\ref{fig:lac-operon-combo}): gene i encodes the \emph{lac
Repressor}, which, in the absence of lactose, binds to gene o (the
\emph{operator}). Transcription of structural genes into mRNA is
performed by the RNA polymerase enzyme, which usually binds to gene
p (the \emph{promoter}) and scans the operon from left to right by
transcribing the three structural genes z, y and a into a single
mRNA fragment. When the lac Repressor is bound to gene o, it becomes
an obstacle for the RNA polymerase, and the transcription of the
structural genes is not performed. On the other hand, when lactose
is present inside the bacterium, it binds to the Repressor and this
cannot stop anymore the activity of the RNA polymerase. In this case
the transcription is performed and the three enzymes for lactose
degradation are synthesized.

\subsection{Typed Stochastic CLS Model}

A detailed mathematical model of the regulation process can be found
in \cite{WGK97}. It includes information on the influence of lactose
degradation on the growth of the bacterium.

We give a TSCLS model of the gene regulation process, with
stochastic rates taken from~\cite{Wil06}. We model the membrane of
the bacterium as the looping sequence $\Loop{m}$, where the alphabet
symbol $m$ generically denotes the whole membrane surface in normal
conditions. Moreover, we model the lactose operon as the sequence
$lacI \cdot lacP \cdot lacO \cdot lacZ \cdot lacY \cdot lacA$
($lacI\!\!-\!\!A$ for short), in which each symbol corresponds to a
gene. We replace $lacO$ with $RO$ in the sequence when the lac
Repressor is bound to gene o, and $lacP$ with $PP$ when the RNA
polymerase is bound to gene p. When the lac Repressor and the RNA
polymerase are unbound, they are modelled by the symbols $repr$ and
$polym$, respectively. We model the mRNA of the lac Repressor as the
symbol $Irna$, a molecule of lactose as the symbol $LACT$, and beta
galactosidase, lactose permease and transacetylase enzymes as
symbols $betagal, perm$ and $transac$, respectively. Finally, since
the three structural genes are transcribed into a single mRNA
fragment, we model such mRNA as a single symbol $Rna$.

The transcription of the DNA, the binding of the lac Repressor to
gene o, and the interaction between lactose and the lac Repressor
are modelled by the following set of stochastic typed rewrite rules:

\begin{equation*}\tag*{(R1)}
lacI\!\!-\!\!A \pipe X
\lmap{\g}{\sn{\sn{}}}{} lacI\!\!-\!\!A\pipe Irna \pipe X\end{equation*}
where $\g=0.02$.

\begin{equation*}\tag*{(R2)}
Irna \pipe X
\lmap{\g}{\sn{\sn {t}}}{} Irna\pipe repr \pipe X\end{equation*}
where $t$ is the type of $Irna$ and $\g=\lambda n.(n+1)\times 0.1$.

\begin{equation*}\tag*{(R3)}
lacI\!\!-\!\!A \pipe polym \pipe X
\lmap{\g}{\sn{\sn{t}}}{} lacI \cdot PP \cdot lacO \cdot lacZ \cdot lacY \cdot lacA \pipe X\end{equation*}
where $t$ is the type of $polym$ and $\g=\lambda n.(n+1)\times 0.1$.

\begin{equation*}\tag*{(R4)}
lacI \cdot PP \cdot lacO \cdot lacZ \cdot lacY \cdot lacA \pipe X
\lmap{\g}{\sn{\sn{}}}{}lacI\!\!-\!\!A \pipe polym \pipe X\end{equation*}
where $\g=0.01$.

\begin{equation*}\tag*{(R5)}
lacI \cdot PP \cdot lacO \cdot lacZ \cdot lacY \cdot lacA \pipe X
\lmap{\g}{\sn{\sn{}}}{}lacI\!\!-\!\!A \pipe polym \pipe Rna \pipe X\end{equation*}
where $\g=20$.

\begin{equation*}\tag*{(R6)}
Rna \pipe X
\lmap{\g}{\sn{\sn {t}}}{} Rna\pipe betagal \pipe perm \pipe transac \pipe X\end{equation*}
where $t$ is the type of $Rna$ and $\g=\lambda n.(n+1)\times 0.1$.

\begin{equation*}\tag*{(R7)}
lacI\!\!-\!\!A \pipe repr \pipe X
\lmap{\g}{\sn{\sn{t}}}{} lacI \cdot lacP \cdot RO \cdot lacZ \cdot lacY \cdot lacA \pipe X\end{equation*}
where $t$ is the type of $repr$ and $\g=\lambda n.(n+1)\times 1$.

\begin{equation*}\tag*{(R8)}
lacI \cdot PP \cdot lacO \cdot lacZ \cdot lacY \cdot lacA \pipe repr \pipe X
\lmap{\g}{\sn{\sn t}}{} lacI \cdot PP \cdot RO \cdot lacZ \cdot lacY \cdot lacA  \pipe X\end{equation*}
where $t$ is the type of $repr$ and $\g=\lambda n.(n+1)\times 1$.

\begin{equation*}\tag*{(R9)}
lacI \cdot lacP \cdot RO \cdot lacZ \cdot lacY \cdot lacA \pipe X
\lmap{\g}{\sn{\sn{}}}{}lacI\!\!-\!\!A \pipe repr \pipe X\end{equation*}
where $\g= 0.01$.

\begin{equation*}\tag*{(R10)}
lacI \cdot PP \cdot RO \cdot lacZ \cdot lacY \cdot lacA \pipe X
\lmap{\g}{\sn{\sn{}}}{}lacI \cdot PP \cdot lacO \cdot lacZ \cdot lacY \cdot lacA \pipe repr  \pipe X\end{equation*}
where $\g= 0.01$.

\begin{equation*}\tag*{(R11)}
repr\pipe LACT\pipe X
\lmap{\g}{\sn{\sn {t_r,t_l}}}{} RLACT\pipe X\end{equation*}
where $t_r$ and $t_l$ are the types of $repr$ and $LACT$ and $\g=\lambda n_1n_2.(n_1+1)\times(n_2+1)\times 0.005$.

\begin{equation*}\tag*{(R12)}
RLACT\pipe X
\lmap{\g}{\sn{\sn {t}}}{} repr\pipe LACT\pipe X\end{equation*}
where $t$ is the type of $RLACT$ and $\g=\lambda n.(n+1)\times 0.1$.

Rules (R1) and (R2) describe the transcription and translation of
gene i into the lac Repressor (assumed for simplicity to be
performed without the intervention of the RNA polymerase). Rules
(R3) and (R4) describe binding and unbinding of the RNA polymerase
to gene p. Rules (R5) and (R6) describe the transcription and
translation of the three structural genes. Transcription of such
genes can be performed only when the sequence contains $lacO$
instead of $RO$, that is when the lac Repressor is not bound to gene
o. Rules (R7)-(R10) describe binding and unbinding of the lac
Repressor to gene o. Finally, rules (R11) and (R12) describe the
binding and unbinding, respectively, of the lactose to the lac
Repressor.

The following rules describe the behaviour of the three enzymes for
lactose degradation:

\begin{equation*}\tag*{(R13)}
\Loop{\xx} \into (perm \pipe X)\pipe Y\lmap{\g}{\sn{\sn{},\sn{t},\sn{}}}
\Loop{perm \mycdot \xx} \into X \pipe Y\end{equation*}
where $t$ is the type of $perm$ and $\g=\lambda n.(n+1)\times 0.1$.

\begin{equation*}\tag*{(R14)}
\Loop{\xx} \into X\pipe LACT \pipe Y\lmap{\g}{\sn{\sn{\st{t_p}},\sn{},\sn{t_l}}}
\Loop{\xx} \into (LACT \pipe X)\pipe Y\end{equation*}
where $t_p$ and $t_l$ are the types of $perm$ and $LACT$, respectively, and $\g=\lambda n_1n_2.n_1 \times (n_2+1)\times 0.001$.

\begin{equation*}\tag*{(R15)}
LACT\pipe X
\lmap{\g}{\sn{\sn {t_l,t_b}}}{} GLU \pipe GAL \pipe X\end{equation*}
where $t_l$ and $t_b$ are the types of $LACT$ and $betagal$,  and $\g=\lambda n_1n_2. (n_1+1)\times n_2 \times 0.001$.

Rule (R13) describes the incorporation of the lactose permease in
the membrane of the bacterium, rule (R14) the transportation of
lactose from the environment to the interior performed by the
lactose permease, and rule (R15) the decomposition of the lactose
into glucose (denoted GLU) and galactose (denoted GAL) performed by
the beta galactosidase.

The initial state of the bacterium when no lactose is present in the
environment and when 100 molecules of lactose are present are
modelled, respectively, by the following terms (where $n \times T$
stands for a parallel composition $T\pipe \ldots \pipe T$ of length
$n$):
\begin{gather}
Ecoli \qqop{::=} \Loop{m} \into (lacI\!\!-\!\!A \pipe 30 \times
polym \pipe 100
\times repr)\label{eq:ecoli}\\
EcoliLact \qqop{::=} Ecoli \pipe 100 \times LACT
\label{eq:ecolilact}
\end{gather}
\normalsize

Now, starting from the term $EcoliLact$, a possible stochastic trace generated by our semantics, given the rules above, is\footnote{For simplicity we just show the rate of the transition reaching the target state considered in the trace. We avoid to report explicitly the whole exit rate from a given term, which should be computed, following the standard simulation algorithm, by summing up the rates for all the possible target states. For the sake of readability, we also show, on the transitions, the labels of the rules leading the state change.}:

\[EcoliLact\ltrans{\textrm{R3},\;  30 \times 0.1} 100 \times LACT \pipe \Loop{m} \into (lacI \cdot PP \cdot lacO \cdot lacZ \cdot lacY \cdot lacA \pipe 29 \times
polym \pipe 100\times repr)\]
\[\ltrans{\textrm{R5},\; 20} 100 \times LACT \pipe \Loop{m} \into (lacI\!\!-\!\!A \pipe 30 \times
polym \pipe 100 \times repr\pipe Rna)\]
\[ \ltrans{\textrm{R6},\;  0.1} 100 \times LACT \pipe \Loop{m} \into (lacI\!\!-\!\!A \pipe 30 \times
polym \pipe 100\times repr \pipe Rna \pipe betagal \pipe perm \pipe transac)\]
\[ \ltrans{\textrm{R13}, \; 0.1} 100 \times LACT \pipe \Loop{perm \mycdot m} \into (lacI\!\!-\!\!A \pipe 30 \times
polym \pipe 100\times repr \pipe Rna \pipe betagal \pipe transac)\]
\[ \ltrans{\textrm{R14},\;  100 \times 0.001} 99 \times LACT \pipe \Loop{perm \mycdot m} \into (lacI\!\!-\!\!A \pipe 30 \times
polym \pipe 100\times repr \pipe Rna \pipe betagal \pipe transac \pipe LACT)\]
\[ \ltrans{\textrm{R15}, \; 0.001} 99 \times LACT \pipe \Loop{perm \mycdot m} \into (lacI\!\!-\!\!A \pipe 30 \times
polym \pipe 100 \times repr\pipe Rna \pipe betagal \pipe transac \pipe GLU \pipe GAL)\]

\section{Conclusions and Related Work}\label{secRelatedWorks}

This paper is a first proposal for using types in describing
quantitative aspects of biological systems. Types for qualitative
properties of the CSL calculus have been studied in \cite{ADT08} and
\cite{DGT09}. We plan to develop a prototype simulator for our
calculus TSCLS in order to experimentally test the expressiveness of
our formalism. This would make possible to compare quantitatively
the approach presented in this paper, with the one of
\cite{BMMTT08}.

In the remaining of this section we will put our paper in the
framework of qualitative and quantitative models of  biological
systems.

\paragraph{Qualitative Models.}

In the last few years many formalisms originally developed by
computer scientists to model systems of interacting components have
been applied to Biology. Among these, there are Petri
Nets~\cite{MDNM00}, Hybrid Systems~\cite{ABI01}, and the
$\pi$-calculus~\cite{CDPB04,RSS01}. Moreover, new formalisms have
been defined for describing biomolecular and membrane
interactions~\cite{BMMT06,Car05,CCD04,DL04,PQ05,RPSCS04}. Others,
such as P-Systems \cite{P02}, have been proposed as biologically
inspired computational models and have been later applied to the
description of biological systems.

The $\pi$-calculus and new calculi based on it \cite{PQ05,RPSCS04}
have been particularly successful in the description of biological
systems, as they allow describing systems in a compositional manner.
Interactions of biological components are modelled as communications
on channels whose names can be passed; sharing names of private
channels allows describing biological compartments.

These calculi offer very low-level interaction primitives, but may
cause the description models to become very large and difficult to
read. Calculi such as those proposed in \cite{Car05,CCD04,DL04} give
a more abstract description of systems and offer special
biologically motivated operators. However, they are often
specialized to the description of some particular kinds of phenomena
such as membrane interactions or protein interactions.

P-Systems \cite{P02} have a simple notation and are not specialized
to the description of a particular class of systems, but they are
still not completely general. For instance, it is possible to
describe biological membranes and the movement of molecules across
membranes, and there are some variants able to describe also more
complex membrane activities. However, the formalism is not so
flexible to allow describing easily new activities observed on
membranes without extending the formalism to model such activities.

Danos and Laneve~\cite{DL04} proposed the $\kappa$-calculus. This
formalism is based on graph rewriting where the behaviour of
processes (compounds) and of set of processes (solutions) is given
by a set of rewrite rules which account for, e.g., activation,
synthesis and complexation by explicitly modelling the binding sites
of a protein.

The Calculus of Looping Sequences~\cite{BMMT06} has no explicit way
to model protein domains (however they can be encoded, and a variant
with explicit binding has been defined in~\cite{BMM07}), but
accounts for an explicit mechanism (the \emph{looping sequences}) to
deal with compartments and membranes. Thus, while the
$\kappa$-calculus seems more suitable to model protein interactions,
CLS allows for a more natural description of membrane interactions.
Another feature lacking in other formalisms is the capacity to
express ordered sequences of elements. To the best of our knowledge,
CLS is the first formalism offering such a feature in an explicit
way, thus allowing to naturally operate over proteins or DNA
fragments which should be frequently defined as ordered sequences of
elements.

\paragraph{Stochastic Models.}

Among stochastic process algebras we would like to mention the
stochastic extension of the $\pi$-calculus, given by Priami et al.
in~\cite{PRSS01}, and the PEPA framework proposed by Hillston
in~\cite{Hil96}. We also would like to compare our work with two
closer ones, namely~\cite{BMMTT08} and~\cite{BV09}.

The stochastic engine behind PEPA and the Stochastic $\pi$-calculus
is constructed on the intuition of cooperating agents under
different bandwidth limits. If two agents are interacting, the time
spent for a communication is given by the slowest of the agents
involved. Differently, our stochastic semantics is defined in terms
of the collision-based paradigm introduced by Gillespie. A similar
approach is taken in the quantitative variant of the
$\kappa$-calculus (\cite{DFFK07}) and in BioSPi (\cite{PRSS01}).
Motivated by the law of mass action, here we need to count the
number of the reactants present in a system in order to compute the
exact rate of a reaction. In~\cite{KMT08}, a stochastic semantics
for bigraphs has been developed. An application in the field of
systems biology has been provided by modelling a process of membrane
budding.

A stochastic semantics for CLS (SCLS) has been defined in
\cite{BMMTT08}. Such a semantics computes the transition rates by
resorting to a complete counting mechanism to detect all the
possible occurrences of patterns within a term. In our framework,
the set of rule schemata that can be defined is limited with respect
to SCLS, however, our counting mechanism, based on types, is quite
more simple in practice. This would simplify, for example, the
development of automatic simulators. As another advantage, our
rules, similar to what happens in~\cite{BV09} for a variant of the
ambient calculus, are equipped with rate functions, rather than with
rate constants. Such functions may allow the definition of kinetics
that are more complex than the standard mass-action ones.

Bioambients,~\cite{RPSCS04}, is a calculus in which biological
systems are modelled using a variant of the ambient calculus. In
Bioambients both membranes and elements are modelled by ambients,
and activities by capabilities (enter, exit, expel, etc.). In
\cite{BV09}, Bioambients are extended by allowing the rates
associated with rules to be context dependent. Dependency is
realized by associating to a rule a function which is evaluated when
applying the rule, and depends on the context of the application.
The context contains (as for our stochastic contexts) the state of
the sibling ambients, that is the ambients in parallel in the
innermost enclosing ambient (membrane). The property of the context
used to determine the value of the function is its volume that
synthesizes (with a real number) the elements present in the
context. In Section~\ref{SCLS_types} we sketched the representation
of osmosis in our framework: the same example is presented with all
details in~\cite{BV09}. However, our modelling is more general
allowing to focus more selectively on context, and specifying
functions that may also cause inhibition.

Finally MGS, {\tt http://mgs.spatial-computing.org/}, is a domain
specific language for simulation of biological processes. The state
of a dynamical system is represented by a collection. The elements
in the collection represent either entities (a subsystem or an
atomic part of the dynamical system) or messages (signal, command,
information, action, etc.) addressed to an entity. The dynamics is
defined by rewrite rules specifying the collection to be substituted
through a pattern language based on the neighborhood relationship
induced by the topology of the collection. It is possible to specify
stochastic rewrite strategies. In \cite{MSG09}, this feature is used
to provide the description of various models of the genetic switch
of the $\lambda$ phage, from a very simple biochemical description
of the process to an individual-based model on a Delaunay graph
topology. Note that, in MSG, the topological changes are programmed
in some external language, whereas in CLS they are specified
directly by the rewrite rules.

\paragraph{Acknowledgements.} We thank the referees for their helpful comments.
The final version of the paper improved due to their suggestions.

%%%%%%%%%%%%%%%%%%%%%%%%%%%%%%%%%%%%%%%%%%%%%%%%%%%%%%%%%%%%%%%%%%%%%%%%%
%%%%%%%%%%%%%%%%%%%%%%%%%%%%%%%%%%%%%%%%%%%%%%%%%%%%%%%%%%%%%%%%%%%%%%%%%
%%%%%%%%%%%%%%     B  I  B  L  I  O  G  R  A  F  I  A    %%%%%%%%%%%%%%%%
%%%%%%%%%%%%%%%%%%%%%%%%%%%%%%%%%%%%%%%%%%%%%%%%%%%%%%%%%%%%%%%%%%%%%%%%%
%%%%%%%%%%%%%%%%%%%%%%%%%%%%%%%%%%%%%%%%%%%%%%%%%%%%%%%%%%%%%%%%%%%%%%%%%

\end{document}